\documentclass{iau}

\usepackage{aas_macros}
\usepackage{amsmath}
\usepackage{graphicx}
\usepackage{multirow}
\usepackage{sidecap}

\renewcommand*{\bibfont}{\small}

\begin{document}

\lefttitle{Jin et al.}
\righttitle{Exploring the Dynamics of CME-Driven Shocks}

\jnlPage{1}{7}
\jnlDoiYr{2024}
\doival{10.1017/xxxxx}
\volno{388}
\pubYr{2024}
\journaltitle{Solar and Stellar Coronal Mass Ejections}

\aopheadtitle{Proceedings of the IAU Symposium}
\editors{N. Gopalswamy,  O. Malandraki, A. Vidotto \&  W. Manchester, eds.}

\title{Exploring the Dynamics of CME-Driven Shocks by Comparing Numerical Modeling and Observations}


\author{Meng Jin$^{1}$, Gang Li$^{2}$, Nariaki Nitta$^{1}$, Wei Liu$^{1,\, 3}$, Vah\'e Petrosian$^{4}$, Ward Manchester$^{5}$, Christina Cohen$^{6}$, Frederic Effenberger$^{3,7}$, Zheyi Ding$^{8}$, Melissa Pesce-Rollins$^{9}$, Nicola Omodei$^{4}$ and Nat Gopalswamy$^{10}$ }
\affiliation{
$^{1}$ Lockheed Martin Solar and Astrophysics Lab (LMSAL), Palo Alto, CA, USA \\ 
$^{2}$ General Linear Space Plasma Lab LLC, Foster City, CA, USA\\ $^{3}$ Bay Area Environmental Research Institute (BAERI), Mountain View, CA, USA \\ 
$^{4}$ Stanford University, Stanford, CA, USA  \\
$^{5}$ University of Michigan Ann Arbor, Ann Arbor, MI, USA \\
$^{6}$ California Institute of Technology, Pasadena, CA, USA\\
$^{7}$ Ruhr University Bochum (RUB), Bochum, Germany\\
$^{8}$ KU Leuven, Leuven, Belgium\\
$^{9}$ Istituto Nazionale di Fisica Nucleare (INFN), Sezione di Pisa, Pisa, Italy\\
$^{10}$ NASA Goddard Space Flight Center, Greenbelt, MD, USA
}

\begin{abstract}
Shocks driven by coronal mass ejections (CMEs) are primary drivers of gradual solar energetic particle (SEP) events, posing significant risks to space technology and astronauts. Concurrently, particles accelerated at these shocks may also propagate back to the Sun, potentially generating gamma-ray emissions through pion decay. We incorporated advanced modeling and multi-messenger observations to explore the role of CME-driven shocks in gamma-ray emissions and SEPs. Motivated by Fermi-LAT long-duration solar flares, we used the AWSoM MHD model to investigate the connection between the shocks and the properties of observed gamma-ray emissions. By coupling the AWSoM with iPATH model, we evaluate the impact of shock evolution complexity near the Sun on SEP intensity and spectra. Our result points to the importance of accurate background coronal and solar wind modeling, as well as detailed observations of CME source regions, in advancing our understanding of CME-driven shocks and the dynamics of associated energetic particles.


\end{abstract}


\maketitle

\section{Introduction}
Shocks driven by Coronal Mass Ejections (CMEs) play an important role in accelerating particles. Particles cross the shock front back and forth due to interaction with magnetic turbulence and gain energy through the diffusive shock acceleration mechanism \citep[e.g.][]{axford77}. Accelerated particles that escape upstream can propagate along the interplanetary magnetic field lines and are detected in situ as Solar Energetic Particles (SEPs), posing significant risks to both technology and astronauts in space.

Conversely, particles accelerated at the shock front can also escape downstream, traveling back to the Sun where they may generate $\gamma$-ray emissions through pion decay with the ambient plasma in the photosphere. The sustained/long-duration $\gamma$-ray emission (SGRE) events observed by the Fermi Large Area Telescope \citep[LAT:][]{Atwood2009} provide evidence supporting this ``shock scenario", which potentially opens a new window for observing shock-accelerated particles in $\gamma$-rays. \citet{gopal18} found a strong correlation between the SGRE events and Type II radio bursts, with a linear relationship between their durations, suggesting a common shock accelerating both protons and electrons that are respectively responsible for such emissions. In a recent study, \citet{gopal21} found that the number of 500 MeV protons from the SGRE events is correlated with the protons propagating into space as SEPs.

In both cases, the CME-driven shock and its dynamical evolution through the inhomogeneous solar wind are crucial for understanding the particle acceleration and transport processes involved. Employing data-driven global magnetohydrodynamics (MHD) models enables the quantitative reconstruction of the global corona and the dynamic evolution of CME-driven shocks, offering unique insights into the role of CMEs in these phenomena \citep[e.g.][]{plotnikov2017, jin18}. In addition, to properly model SEP events, the MHD models need to be coupled with particle acceleration and transport models \citep[e.g.][]{li21}. This study aims to investigate both $\gamma$-ray and SEP events to delve into the dynamics of CME-driven shocks by comparing advanced modeling and observations results. 

The paper is structured as follows: Section 2 provides an overview of the models used; Section 3 presents the main results; discussions are in Section 4, followed by the summary and conclusions in Section 5.

\section{Model}
To reconstruct the background solar corona and wind, we use the Alfv\'{e}n Wave Solar Model \citep[AWSoM;][]{bart14}, a data-driven global MHD model within the Space Weather Modeling Framework \citep{toth12}. By incorporating a physically consistent treatment of wave reflection, dissipation, and heat partitioning between electrons and protons, AWSoM has been demonstrated to reproduce high-fidelity global solar corona conditions \citep[e.g.,][]{sokolov13, jin17a, Sachdeva19}. Specifically, the inner boundary condition of the magnetic field is specified with full-surface magnetic maps. Aflv\'{e}n waves are driven at the inner boundary with a Poynting flux that scales with the surface magnetic field. The solar wind is heated by Alfv\'{e}n wave dissipation and accelerated by thermal and Aflv\'{e}n wave pressure. While electrons and protons are collisionally coupled, their temperatures are computed separately. This treatment is critical for reproducing physically correct CME-driven shocks \citep{jin13}.

To generate a CME eruption, we use a simulation tool called the Eruptive Event Generator Gibson-Low \citep[EEGGL, now available at NASA CCMC,][]{jin17b, borovikov17} that inserts a Gibson-Low (GL) flux rope \citep{gibson98} into the corona. It automatically determines the flux rope parameters with a handful of quantities given by observations (i.e., magnetogram and CME speeds) so that the modeled CMEs can travel at the desired speeds near the Sun.

To model SEP events, we further couple the AWSoM MHD model with the improved Particle Acceleration and Transport in the Heliosphere  model \citep[iPATH:][]{hu2017}. iPATH numerically solves the particle acceleration at the shock and transport in the solar wind with both along- and cross-field diffusion. One unique feature of iPATH is the adoption of a 2D onion shell module which tracks the acceleration of particles and their subsequent convection and diffusion downstream of the shock. \citet{li21} extended the 2D shell structures to 3D using the AWSoM MHD shock and plasma inputs. In addition, an ensemble approach was developed to account for the model uncertainties in magnetic connectivity and shock profiles, i.e., multiple field lines were traced that were evenly distributed within 10$^{\circ}$ of a heliospheric observer.

\section{Results}
\subsection{CME-driven Shock in Relation to Fermi SGRE Solar Flares}

In this section, we show the modeling results for three Fermi SGRE events. Two of them (SOL2014-09-01 and SOL2021-07-17) are behind-the-limb (BTL) flares \citep{Pesce-Rollins2015, Ackermann2017}, in which the $\gamma$-ray emission region on the visible side of the Sun was located \emph{away} from the flare site behind the limb by tens of degrees in longitude. The other, equally intriguing, event is SOL2012-03-07, in which the Fermi $\gamma$-ray emission centroid \emph{migrated with time} over the solar disk for 10~hours, long past the impulsive phase of the flare \citep{Ajello2021}. 

\begin{figure}[htb]
  \includegraphics[width=1.0\textwidth]{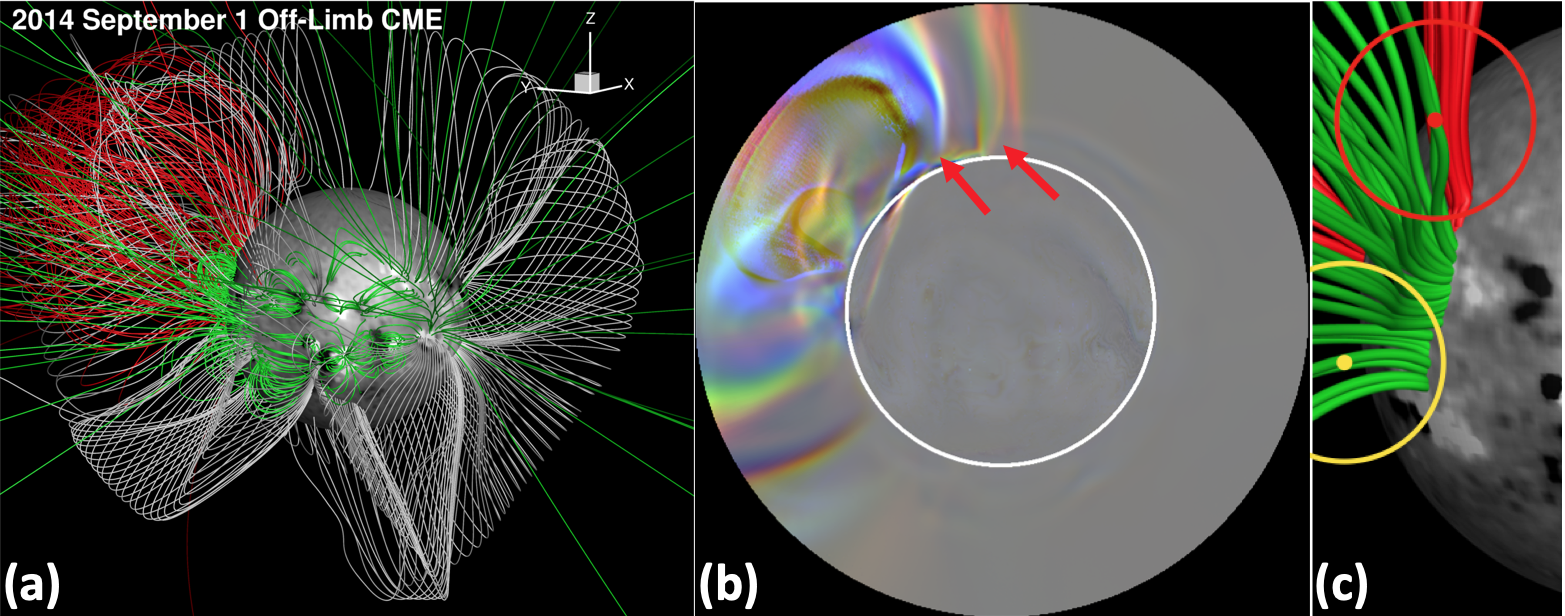}
    \caption{MHD modeling result of the SOL2014-09-01 Fermi BTL event. (a) Magnetic field configuration at $t = 16$~minutes after the flux rope eruption. The red and white field lines represent the flux rope and large-scale helmet streamers, respectively. The green field lines are selected surrounding active regions as well as open field lines. (b)  Concurrent synthesized tri-color composite, running-ratio image for AIA 211~\AA (red), 193~\AA (green), and 171~\AA (blue) channels showing the simulated EUV waves. (c) Selected field lines connecting to the CME shock (green) and source flare region (red). The red and yellow dots with 95\% uncertainty circles represent $>$100 MeV emission centroids identified previously by \citet{Ackermann2017} and updated recently in \citet{Ajello2021}, respectively.}
  \label{fig:fermi_140901}
\end{figure}

In the SOL2014-09-01 event, the solar flare region was located 43$^{\circ}$ behind the east limb. However, Fermi-LAT detected $\gamma$-ray emissions on the front side of the Sun for $\sim$2 hours. This event was associated with a fast CME with a speed $>$1900 km s$^{-1}$. Figure~\ref{fig:fermi_140901} shows the MHD modeling result of this event. Specifically, Figure~\ref{fig:fermi_140901}a depicts the magnetic field configuration after 16 minutes of the CME onset, while Figure~\ref{fig:fermi_140901}b illustrates the simulated global EUV waves associated with the CME, where two different wave fronts are evident (marked in red arrows). The first wave front (to the right) is the fast-mode wave/shock driven by the eruption. The second one (to the left) is caused by the expanding CME flux rope shell \citep[e.g.,][]{jin16}. This double wave front feature is consistent with SDO/AIA observations of the event \citep[e.g.,][]{grechnev18}. Note that EUV waves are thought to be the low coronal counterparts of the CME-driven shock waves that propagate into the heliosphere. A recent study by \citet{Pesce-Rollins2022} demonstrated a strong correlation between the time derivative of the EUV wave intensity profile and Fermi/LAT $\gamma$-ray flux.

By modeling the evolution of the CME and the CME-driven shock over one hour, we derived the detailed history of 3D shock parameters including compression ratio, shock speed, shock Alfv\'{e}n Mach number, and shock angle. We found that the shock compression ratio increases rapidly from $\sim$1.8 at 10 minutes to $\sim$4.6 at 20 minutes and then gradually decreases to $\sim$3.7 at 60 minutes. This evolution trend is similar to the Fermi/LAT $\gamma$-ray intensity profile. Figure~\ref{fig:fermi_140901}c shows the field lines connected to the CME shock (in green) and the source region (in red). Notably, in the recently updated Fermi solar flare catalog, the $>$100 MeV emission centroid has been revised further south (marked by a yellow circle), thus aligning more closely with the shock-connected field lines. See \citet{jin18} for detailed modeling results of this event.

\begin{figure}
  \includegraphics[width=1.0\textwidth]{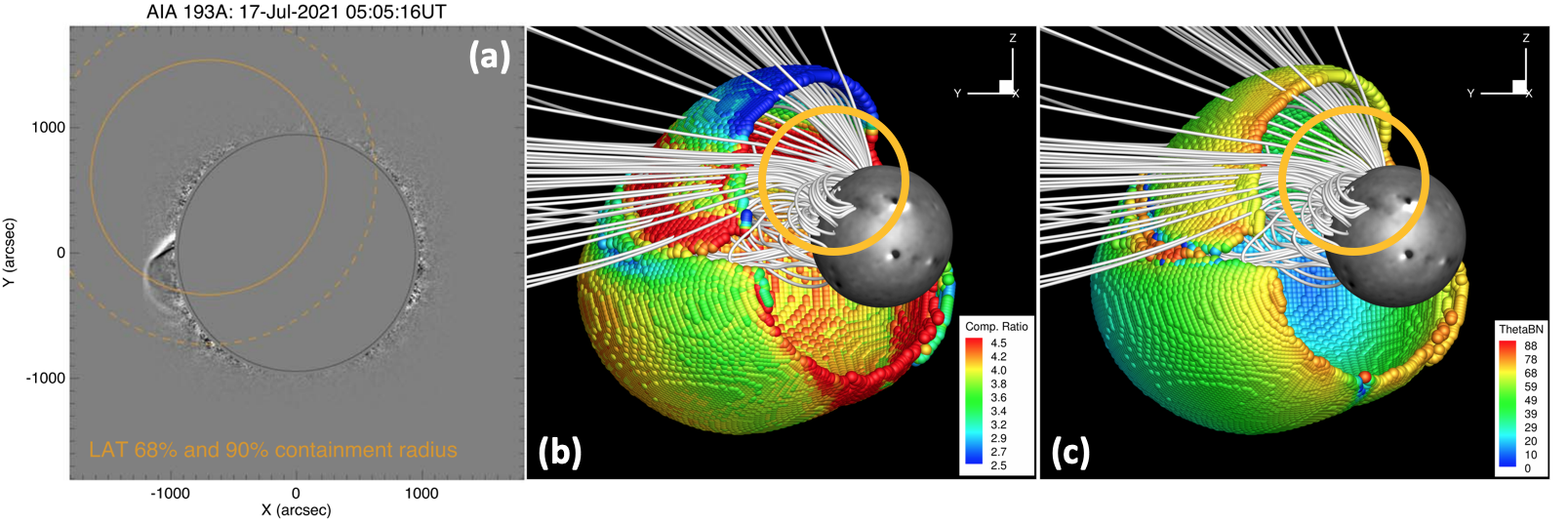}
    \caption{(a) Flare averaged Fermi-LAT emission centroid, shown as 90\% (solid) and 68\% (dashed) uncertainty circles, overlaid on the SDO/AIA 193~\AA~running difference image \citep[adpated from ][]{Pesce-Rollins2022}. (b)-(c) Modeled CME-driven shock compression ratio and $\theta_{Bn}$ at t = 20 minutes after eruption. The orange circle shows the same 90\% uncertainty radius of the Fermi-LAT emission centroid as in (a).}
  \label{fig:fermi_210717}
\end{figure}

SOL2021-07-17 is the most recent Fermi BTL event observed (Figure~\ref{fig:fermi_210717}), and it exhibits the largest separation ($>$50$^{\circ}$) between the source region and the emission centroid of all the Fermi BTL events \citep{Pesce-Rollins2022}. Preliminary modeling results of the event are shown in Figure~\ref{fig:fermi_210717}b-c, which include the 3D shock compression ratio and shock angle from the simulation. The shock surface is notably non-spherical due to the inhomogeneous background solar wind, leading to significant variations in the shock parameters across different parts of the shock surface. Similar to the SOL2014-09-01 event, this event features shock-connecting open field lines that trace back to the Fermi emission centroid area. Additionally, the area where the shock intersects these field lines exhibits a quasi-perpendicular nature.

\begin{figure}
  \includegraphics[width=1.0\textwidth]{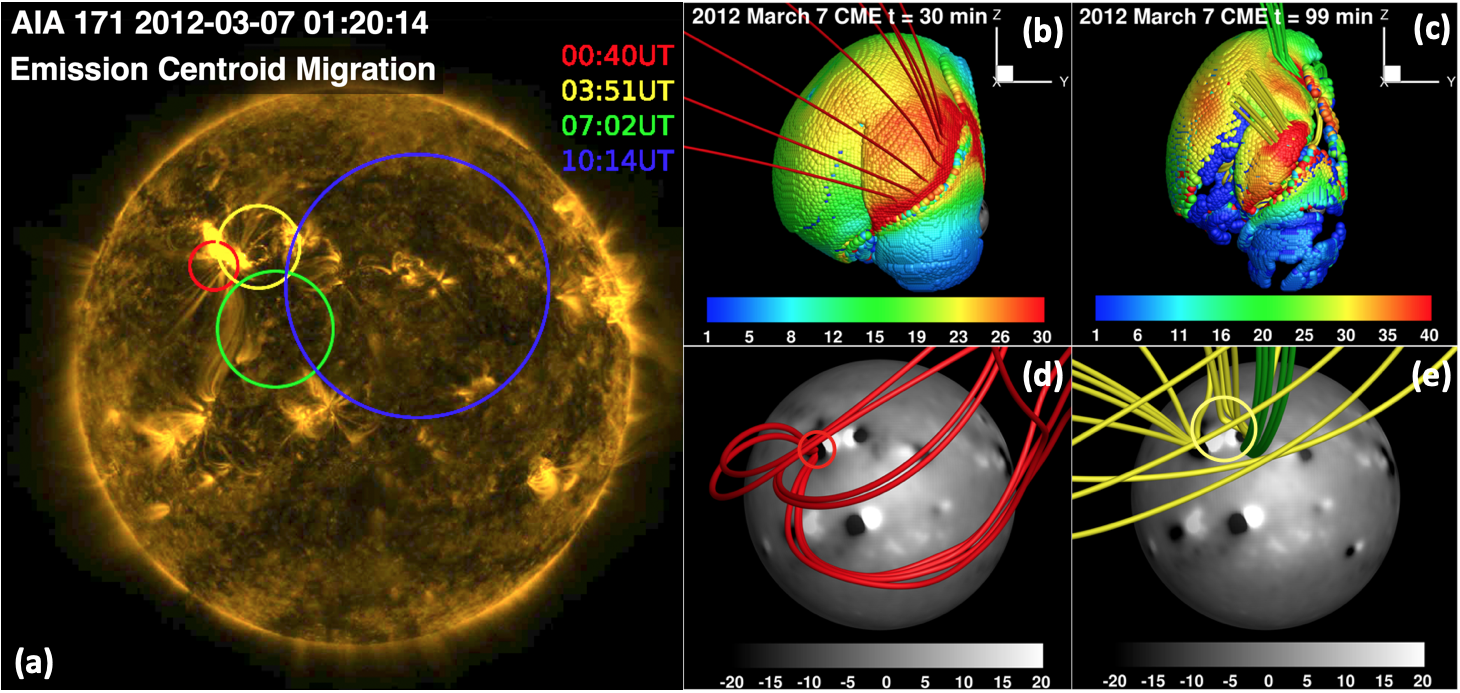}
    \caption{(a) Migration of the $\gamma$-ray emission centroid of the SOL2012-03-07 flare over 10~hours overlaid on an AIA 171~\AA\ image at 00:40~UT. Circles show the 95\% uncertainty. (b) \& (c) Shock surface with the shock strength ($CR \cdot {M}_A$) in color derived from the simulation. (d) \& (e) Magnetic field lines connecting to the strongest portion of the shock. The red and yellow circles are reproduced from (a), indicating good agreement of the LAT centroids with the roots of the field lines in corresponding colors.}
  \label{fig:fermi_120307}
\end{figure}

Lastly, we discuss the SOL2012-03-07 event. In this event Fermi observed the $\gamma$-ray emission centroid migration over the solar disk in 10 hours after flare impulsive phase (Figure~\ref{fig:fermi_120307}a). This event is more challenging to model since there are two CMEs involved, separated by 1 hour. Although the two CMEs are from the same active region, they erupted from different parts of the polarity inversion line (PIL), with the first one started from the north, and the second one from the south \citep[e.g.,][]{patsourakos16}. 

We model the SOL2012-03-07 event by initiating two separate GL flux ropes at different parts of PIL to match the AIA observation. To estimate strong shock location in the simulation, we define a new parameter that combines compression ratio and Alfv\'{e}n Mach number as shock strength measure. This approach enabled us to identify areas of strong shock on the shock surface and trace the field lines from those regions back to the Sun's surface (Figure~\ref{fig:fermi_120307}). At 30 minutes, these footpoints are situated south of the active region, aligning with the initial Fermi centroid marked in red. By 99 minutes into the simulation, the footpoints connected to the strong shock location begin moving northward, as indicated by the yellow circle, a movement consistent with observations.

Furthermore, we observed another strong shock region that developed at 99 minutes, with its footpoints situated near the Fermi centroid at 07:02, marked by a green circle. While we continue to work on the simulation beyond 100 minutes, these preliminary results indicate that the temporal evolution of footpoints, those which are connected to the strongest shock locations,  aligns with the migration of the Fermi centroid during the first 100 minutes.

\subsection{The Impact of CME-driven Shock Evolution on the in-situ SEP Intensity and Spectra}
In this section, we discuss the impact of the CME-driven shock on the properties of SEPs measured in situ. As demonstrated in the previous section, the parameters of the CME-driven shock can vary significantly across different parts of the shock surface. Given that heliospheric observers at various locations are connected to different parts of the shock, the characteristics of the SEPs observed can be influenced by the properties of the specific shock segment to which they are connected. Understanding this relationship is crucial for better interpreting SEP observations.

\begin{figure}
  \includegraphics[width=1.0\textwidth]{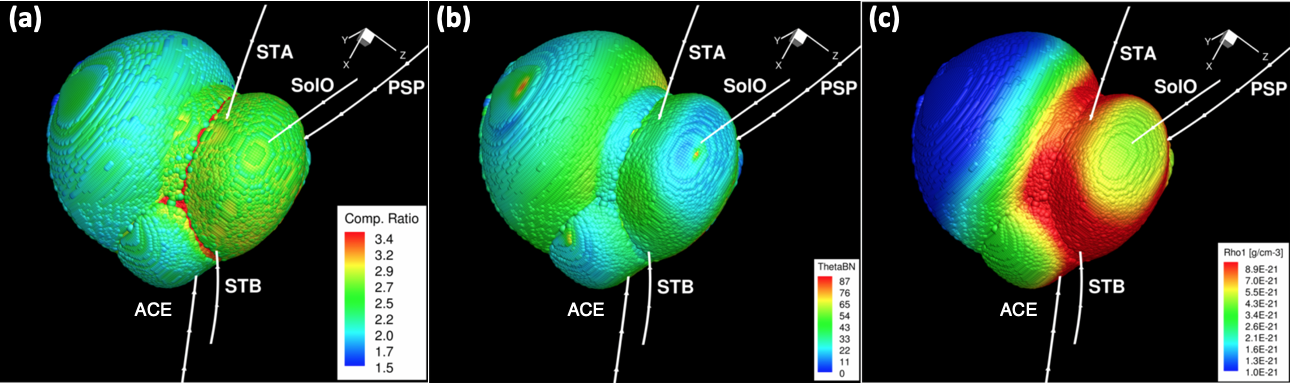}
    \caption{(a) The shock compression ratio, (b) shock angle, and (c) upstream plasma density along the shock surface at t = 1 hour in the 2021 October 28 SEP event. The white lines represent the field connectivity to different heliospheric observers.}
  \label{fig:PSP_211028}
\end{figure}

One example is the 2021 October 28 SEP event, which was observed by Parker Solar Probe (PSP), STEREO-A, and SOHO/ACE with almost identical Fe/O ratio, even there is $\sim$60$^{\circ}$ separation in longitude \citep{cohen22}. To better understand this distinctive characteristic, we modeled the associated CME-driven shock of this SEP event and calculated the field line connectivity to different observer locations including the SOHO/ACE, STA/B, SolO, and PSP (Figure~\ref{fig:PSP_211028}). Our preliminary results indicate that although the three observers (i.e., PSP, STEREO-A, and ACE) are connected to different parts of the shock, these areas nevertheless exhibit very similar properties, e.g., compression ratio, shock angle (see Figure~\ref{fig:PSP_211028}a \& b), as well as the upstream condition (see Figure~\ref{fig:PSP_211028}c for upstream plasma density). This consistency might explain the similar Fe/O ratios observed at these locations.

Fe-rich events in gradual SEP events are thought to originate either directly from flare contributions \citep[e.g.,][]{cane03,cane06} or from suprathermal flare particles that are preferentially accelerated by quasi-perpendicular shocks \citep{tylka05}. To evaluate the ``quasi-perpendicular shock scenario", eight SEP events were modeled using AWSoM MHD CME simulations, which allow for direct assessment of shock geometry — a feature difficult to discern from near-Sun observations \citep{nitta22}. However, the preliminary results do not indicate a strong correlation between Fe-rich SEP events and quasi-perpendicular shocks. Suprathermal particles, which can be affected significantly by the inhomogeneity of the solar wind \citep{Wijsen2023}, or due to preceding impulsive SEP events, may play a significant role and warrant further investigation.

\begin{figure}
  \centering
  \includegraphics[width=0.8\textwidth]{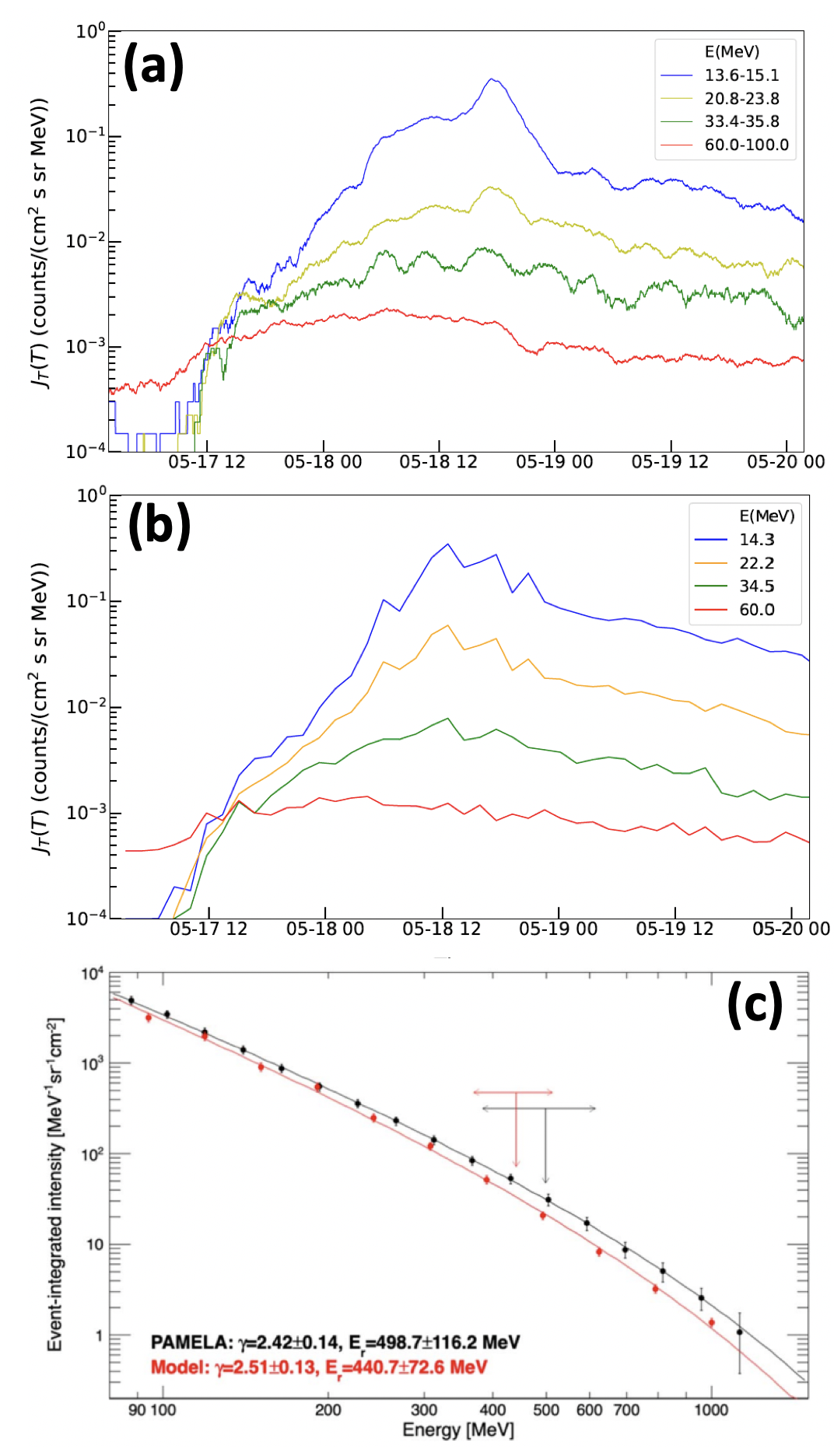}
    \caption{(a) Proton time-intensity profiles for 4 energy bins observed at STEREO-A during the 2012 May 17 SEP event. (b) Modeled proton time-intensity profiles from AWSoM+iPATH. (c) Spectrum comparison between PAMELA observation and model simulation. The black and red arrows indicated the location of the roll-over energies. Figures adapted from \citet{li21}.}
  \label{fig:SEPCaster}
\end{figure}

To quantitatively investigate the SEP events and their relationship to the CME-driven shocks, one needs both the global MHD model for tracking the CME-driven shock evolution through the background solar wind and the particle acceleration and transport model for calculating the transport of particles along the magnetic field lines. Most SEP models rely on simplified background solar wind and start at a larger distance from the Sun (e.g., 10 Rs). However, CME-driven shock could form at a height of just a few solar radii, with information on timing and formation height revealed through radio observations \citep[e.g.,][]{kozarev11, gopal13}. There are previous attempts to include early phase shock evolution \citep[e.g.,][]{kozarev13, borovikov18, linker19}. By coupling two state-of-the-art models AWSoM and iPATH, \cite{li21} modeled the 2012 May 17 SEP event, including the very early phase. The new model successfully reproduced the SEP intensity profiles at both the Earth and STEREO-A locations as shown in Figure~\ref{fig:SEPCaster}a-b. The new model also reproduces the high-energy end of SEP spectra up to 1 GeV observed by PAMELA \citep{bruno18}. The roll-over feature at high energy can be seen in both the observation and the simulation (Figure~\ref{fig:SEPCaster}c). This study emphasizes the importance of early phase shock evolution on the particle acceleration and transport, especially on the high-energy end of SEP spectra. A new SEP forecasting model (SEPCaster) is under development based on this new capability \citep{whitman23}. 

\section{Discussion}
Regarding the Fermi-LAT SGRE solar flares, our MHD simulations have shown some promising results supporting the ``shock scenario". However, recent observational studies challenge the notion that the shock is the origin of the $\gamma$-ray producing particles \citep[e.g.,][]{denolfo18, bruno23}. \citet{denolfo18} analyzed the particles interacting with the Sun as observed by Fermi/LAT and the SEPs observed at 1 AU from PAMELA in 14 SGRE solar flare events. Their findings suggest that the two particle populations are not correlated. \citet{gopal21} offered a possible explanation by addressing two observational effects: the $\gamma$-ray flux might be underestimated in limb events, and the latitudinal widths of SEP distributions are energy-dependent.

There are alternative scenarios coexisting with the ``shock scenario". For example, flare-accelerated particles that are trapped locally within extended coronal loops after an eruption could be the source of the $\gamma$-ray producing particles \citep[e.g.,][]{ryan91, hudson18}. In the simulation of the SOL2014-09-01 event, we found some large-scale closed loops connected to the source region due to the interaction between the flux rope magnetic field and the global solar corona. These field lines could be significant for examining these trapped particles. In summary, further information remains to be explored in the simulations, and more advanced modeling is necessary to further differentiate between the scenarios mentioned above.

Another challenge to the ``shock scenario" is the effect of magnetic mirroring. The strong magnetic mirroring, resulting from a high degree of convergence of magnetic field lines towards the Sun and extremely small loss cones, could potentially prevent particles from reaching the photosphere to produce $\gamma$-rays. To assess the impact of magnetic mirroring that inhibits particle transport back to the Sun, we calculated the evolution of the mirror ratio in the SOL2014-09-01 simulation. Using an approximate relationship between the escape time and mirror ratio \citep{malyshkin01, effenberger18}, we determined that the particle escape timescale is shorter than the $\gamma$-ray emission period, which means a significant portion of the downstream protons can reach the photosphere and produce $\gamma$-rays in this event. Future studies should self-consistently calculate the transport of back-propagating protons downstream of the CME-driven shock by combining theory and MHD modeling results \citep[e.g.,][]{petrosian23}.

\begin{figure}[htb]
  \includegraphics[width=1.0\textwidth]{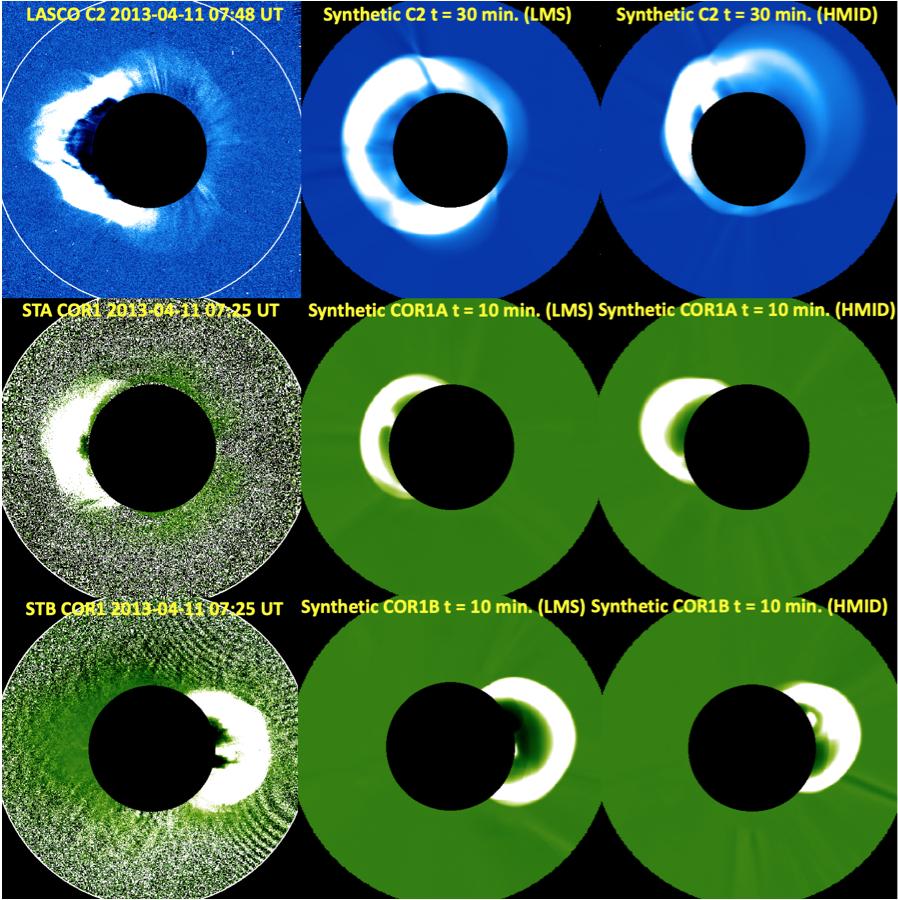}
    \caption{Left: White-light observations from LASCO C2, STEREO-A COR1 and STEREO-B COR1 of 2013 April 11 CME. Middle: Synthesized white light images from the simulation using Lockheed Martin Synchronic (LMS) map. Right: Synthesized white light images from the simulation using HMI Diachronic (HMID) map.}
  \label{fig:cme_130411}
\end{figure}

Finally, we would like to discuss one of the major challenges in CME-driven shock and SEP modeling: the influence of the background solar wind on the properties of the CME-driven shock \citep{jin22}. There are previous studies \citep[e.g.,][]{Hinterreiter19} show the uncertainties of the background solar wind modeling. Figure~\ref{fig:cme_130411} illustrates how the ambient solar corona and wind can significantly affect the parameters of the CME-driven shock. The figure shows the synthesized CME white light images driven by an identical flux rope, but under slightly different background solar wind conditions caused by different types of input magnetic maps (one is a synchronic map, and the other is a diachronic/synoptic map). These variations in the ambient solar wind lead to distinctly different CME-driven shocks. The morphology of the shock surface shows greater latitudinal expansion in the Lockheed Martin
Synchronic (LMS) case than in the HMI Diachronic (HMID) case. The resulting shock parameters are also noticeably different. For detailed comparison of the two cases, we refer to \citet{jin22}. This result underscores the importance of obtaining better constraints on the plasma environment of the CME source region and its path, especially for accurately capturing the early phase of shock evolution.

\section{Conclusions}
The CME-driven shock plays a crucial role in accelerating particles in the heliosphere. By integrating advanced modeling with multi-messenger observations, we have demonstrated the significant impact CME-driven shocks have on both $\gamma$-ray emissions and in-situ SEP observations. Our results provide promising evidence that particles accelerated by CME shocks could travel back to the Sun, producing $\gamma$-rays observed by Fermi and potentially opening a new avenue for observing shock-accelerated particles. However, alternative scenarios also exist, and more detailed modeling and observational efforts are required to enhance our understanding of the physical mechanisms behind Fermi SGRE events. By coupling AWSoM with iPATH and involving the early-stage ($<$10 Rs) shock evolution, our new model shows promising potential to reproduce SEP intensity profiles and spectra at various locations in the heliosphere. Lastly, our findings underscore the importance of accurate background coronal and solar wind modeling, as well as detailed observations of CME source regions, in advancing our understanding of CME-driven shocks and the dynamics of associated energetic particles.

\acknowledgements{M.J., N.N., W.L. acknowledge support from NASA’s SDO/AIA contract (NNG04EA00C) to LMSAL. AIA is an instrument onboard the Solar Dynamics Observatory, a mission for NASA’s Living With a Star program. The authors also acknowledge support from multiple NASA grants 80NSSC24K0136, 80NSSC21K1782, 80NSSC21K1327, 80NSSC18K1126, and 80NSSC24K0175.}


\end{document}